\documentclass[pra,prl,superscriptaddress,floatfix,showpacs,12pt,letterpaper,nofootinbib]{revtex4}
\usepackage{hyperref}
\usepackage{graphicx}
\usepackage{color}
\usepackage{amsmath}
\usepackage{colordvi}

\newcommand{\Tiling}[2][]{%
\edef\Temp{#1}%
\begin{pspicture}#2
\ifx\Temp\empty \psframe[fillstyle=boxfill]#2 \else
\psframe[fillstyle=boxfill,#1]#2 \fi
\end{pspicture}}

\def\ket#1{|#1\rangle}

\def\TOFFOLI(#1,#2,#3,#4){\multCNOT[#1,#2](#3,#4)}
\def\NOT(#1,#2){\OneQubitGate(#1,#2){$X$}}
\def\CZGate(#1,#2,#3){\CGate(#1,#2,#3){$Z$}}
\begin{document}
\title{Experimental Implementation of a Codeword Stabilized Quantum Code}

\author{Jingfu Zhang}
\affiliation{Institute for Quantum Computing and Department of
Physics, University of Waterloo, Waterloo, Ontario, N2L 3G1, Canada}

\author{Markus Grassl}
\affiliation{Centre for Quantum
  Technologies, National University of Singapore, Singapore 117543, Singapore}

\author{Bei Zeng}
\affiliation{Department of Mathematics $\&$
  Statistics, University of Guelph, Guelph, Ontario, N1G 2W1, Canada}
\affiliation{Institute for Quantum Computing and Department of
  Combinatorics $\&$ Optimization, University of Waterloo, Waterloo,
  Ontario, N2L 3G1, Canada}

\author{Raymond Laflamme}
\affiliation{Institute for Quantum Computing and Department of
Physics, University of Waterloo, Waterloo, Ontario, N2L 3G1, Canada}
\affiliation{Perimeter Institute for Theoretical Physics, Waterloo,
Ontario, N2J 2W9, Canada}

\date{\today}

\begin{abstract}
A five-qubit codeword stabilized quantum code is implemented in a
seven-qubit system using nuclear magnetic resonance (NMR). Our
experiment implements a good nonadditive quantum code which
encodes a larger Hilbert space than any stabilizer code with the
same length and capable of correcting the same kind of errors. The
experimentally measured quantum coherence is shown to be robust
against artificially introduced errors, benchmarking the success
in implementing the quantum error correction code.  Given the
typical decoherence time of the system, our experiment illustrates
the ability of coherent control to implement complex quantum
circuits for demonstrating interesting results in spin qubits for
quantum computing.
\end{abstract}
 \maketitle

\section{ introduction}

Quantum computers are promising to solve certain problems faster
than classical computers \cite{bookChuang}. The power of quantum
computing relies on the coherence of quantum states.  In
implementation, however, the quantum devices are subject to
errors, from inevitable coupling to the uncontrollable
environment, or from other mechanisms such as imperfection in
controlled operations.  The errors damage the coherence, and
consequently can reduce the computational ability of quantum
computers. In order to protect quantum coherence, schemes of
quantum error correction and fault-tolerant quantum computation
have been developed
\cite{bookChuang,knill1998,DB97,Knill05,Got06,gottesman97,GF498}.
Those schemes have greatly improved the long-term prospects for
quantum computation technology.

A quantum error correcting code (QECC) protects a $K$-dimensional
Hilbert space (the code space) by encoding it into an $n$-qubit
system, and is usually denoted by parameters $((n,K,d))$, where
$d$ is called the distance of the code \cite{note1}. This
$n$-qubit system is used in the process of quantum computing and
hence subject to errors.  A code of distance $d$ is capable of
correcting $d-1$ erasure errors (i.e. loss of qubits at up to
$d-1$ known positions) or $t=\lfloor\frac{d-1}{2}\rfloor$
arbitrary errors (i.e. arbitrary errors on $t$ qubits at unknown
positions).  At the end of the computation, the quantum code can
be decoded to recover the quantum state in the original
$K$-dimensional Hilbert space.

In practice, one would always hope for a ``good'' QECC where more
information is protected (larger $K$), while less physical
resources are used (smaller $n$) plus more errors can be corrected
(larger $d$). There are trade-offs among these three parameters,
and one can readily develop upper bounds and lower bounds for the
third parameter if two of them are fixed \cite{GF498}. With
increasing length $n$ of the codes, in most of the cases there is
a gap between these upper and lower bounds. Therefore, given two
fixed parameters among the three, to find a code with the best
possible value of the third parameter is one of the most important
topics in studying the theory of QECC.

Stabilizer codes, also known as additive codes, form an important
class of QECCs, developed independently in \cite{gottesman97,GF498} in
the late $1990$s. The construction of these codes is based on a simple
method using Abelian groups, where the code dimension $K$ of
stabilizer codes is always a power of two, that is, $K=2^k$ for some
integer $k$. Stabilizer codes include the shortest
single-erasure-error-correcting code, the $((4,2^2,2))$ code
\cite{VGW96,GBP97}, and the shortest single-arbitrary-error-correcting
code, the $((5,2^1,3))$ code \cite{RayPRL96,BDS+96}.

However, the restriction that the code dimension $K$ of stabilizer
codes is always a power of two indicates that these codes might not be
optimal in many cases. In 1997, a nonadditive code, i.e., a code
outside the class of stabilizer codes, with parameters $((5,6,2))$ was
constructed \cite{RHW+97}. For the given parameters $n=5$ and $d=2$,
this code outperforms the best possible stabilizer code whose
dimension is $K=2^2=4$, providing the first example that good
nonadditive codes exist.  In 2007, after ten year of respite, other
good nonadditive codes were found, including a family of
single-erasure-error-correcting codes containing a code $((5,5,2))$
\cite{Simple07}, and a $((9,12,3))$ single-arbitrary-error-correcting
code \cite{YCL+08}.  In late 2007, a general framework for
constructing nonadditive codes, namely, the codeword stabilized (CWS)
codes, was developed \cite{CWS07}.  These codes encompass stabilizer
codes as well as all good known nonadditive codes. In addition, the
CWS framework provides a powerful method to construct good nonadditive
QECCs in a systematical way, and many good codes outperforming the
best possible stabilizer codes were found
\cite{CWS07,HTZ+08,GR08a,GR08b,GSS+09,GSZ09,YCO09}.

When implementing general QECCs, one big challenge comes from the
complexity of the quantum circuits, for both encoding and decoding. To
verify and enhance the control ability to implement complex quantum
circuits are hence critical tasks for implementing QECCs in building
scalable quantum computers.  Coherent control of the simplest
single-erasure-error-correcting code, the $((4,2^2,2))$ stabilizer
code, has been demonstrated in optical systems \cite{Pan08}.  Coherent
control of the simplest single-arbitrary-error-correcting code, the
$((5,2^1,3))$ stabilizer code, has been implemented in nuclear
magnetic resonance (NMR) systems \cite{RayPRL01}.

Here we implement a $((5,5,2))$ CWS quantum code to benchmark the
ability of coherence control in NMR systems. Experiments are
performed in a seven-qubit system where five spins are used to
represent the five qubits of the code. This $((5,5,2))$ code is
one of the simplest nonadditive codes which encode a larger
dimension of the Hilbert space than the best possible stabilizer
mentioned above, for the given parameters $n=5$ and $d=2$, the
best possible stabilizer code has dimension $K=2^2=4$.  Our
experiment thus gives the first experimental demonstration of a
nonadditive code that outperforms the best possible stabilizer
codes. Implementing one round of experiment typically requires
approximately $320$--$440$ radio frequency (r.f.) pulses with a
total duration of approximately $0.55$--$0.72$s, which is much
longer than for the $((5,2^1,3))$ experiment with $368$ r.f.{}
pulses amounting to a total duration of $0.38$s \cite{RayPRL01}.
 Given that the decoherence time $T_2$ of our NMR
system ranges from about $0.80$s to $1.1$s, our experiment is
quite challenging with respect to coherence control of the system.
 Surprisingly, although only about $10$--$20$\% signal strength
remains available, there is still clear evidence of quantum
coherence, which is robust against the artificial errors generated
by single-qubit rotations. What is more, we can partially
characterize the errors. Our results demonstrate the ability to
control NMR spin systems for implementing complex quantum circuits
for quantum computing \cite{NMRreview}.

\section{Quantum Error-Correcting Codes}
The $((5,5,2))$ code we have implemented has been first introduced
in~\cite{Simple07}, and shown to be a CWS code in~\cite{CWS07}.
This code uses five qubits to protect an arbitrary state in a
five-dimensional Hilbert space, and is capable of correcting a
single erasure error. In other words, if any of the five qubits is
lost, we can always recover the arbitrary state in the
five-dimensional code space as long as we know which qubit was
lost.  This equivalently means that this code corrects an
arbitrary error on any of the five qubits given the error position
is known.

\begin{figure}[hbt]
\includegraphics{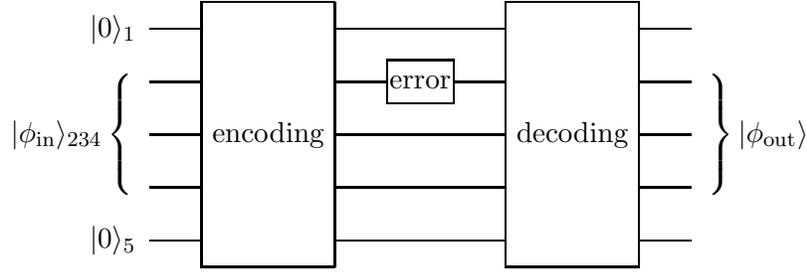}
\caption{Outline of the protocol for the QECC procedure. Five
qubits $1$--$5$ are represented by five lines, from top to bottom,
where qubits $2$, $3$, and $4$ are exploited to carry the input
state $\ket{\phi_{\text{in}}}_{234}$, and qubits $1$ and $5$ are
syndrome qubits. The error can happen on any of the five qubits.
The decoding operation depends on the location of the error. In
the ideal case the output state obeys $\ket{\phi_{\text{out}}} =
\ket{\phi_{\text{in}}}$.
 }\label{figoutline}
\end{figure}

The entire procedure to demonstrate has three parts: encoding the
input state into the five-qubit code, an arbitrary error on one of
the five qubits at a known position, and decoding to recover the
input state.  The outline of the procedure is given in Figure
\ref{figoutline}. We assign qubits $2$--$4$ as register qubits for
carrying the five-dimensional quantum state to be protected, and
choose the five basis states as $\{|b\rangle\}_{b=0}^4$, which in
binary form are
\begin{equation}\label{basis5}
  \{ |000\rangle, |001\rangle, |010\rangle, |011\rangle, |100\rangle\}.
\end{equation}
Then the input state $|\phi_{\text{in}}\rangle_{234}$ can be an arbitrary
superposition of these five basis states.  Qubits $1$ and $5$ are
syndrome qubits, which are initially both in the state
$|0\rangle$.

\begin{figure}[hbt]
\includegraphics{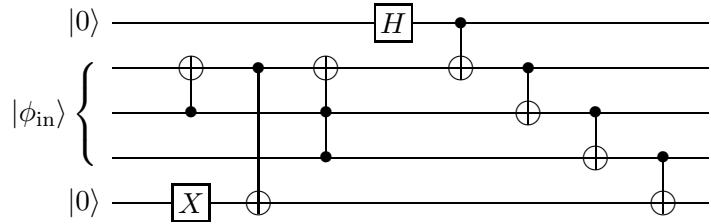}
\medskip
\caption{Quantum circuit for encoding. Elements $\oplus$ and
$\bullet$ connected by a line denote a controlled-NOT or Toffoli
gate conditioned on the state $|1\rangle$. $X$ denotes the Pauli
matrix $\sigma_x$, acting as a NOT gate, and $H$ denotes a Hadmard
gate.}\label{figencode}
\end{figure}

The encoding procedure maps the basis states $\{|b\rangle\}_{b=0}^4$ to
$\{|\varphi_b\rangle\}_{b=0}^4$, which form a basis of the code. Here
 \begin{eqnarray}
 \ket{\varphi_0}&=&\frac{1}{\sqrt{2}}(|00001\rangle+|11110\rangle),\nonumber\\
 \ket{\varphi_1}&=&\frac{1}{\sqrt{2}}(|00010\rangle+|11101\rangle),\nonumber\\
 \ket{\varphi_2}&=&\frac{1}{\sqrt{2}}(|01000\rangle+|10111\rangle),\nonumber\\
 \ket{\varphi_3}&=&\frac{1}{\sqrt{2}}(|00100\rangle+|11011\rangle),\nonumber\\
 \ket{\varphi_4}&=&\frac{1}{\sqrt{2}}(|10000\rangle+|01111\rangle),
 \end{eqnarray}
(see also \cite{Simple07}, but note the re-ordering of the code basis
and the different input states for the encoding circuit).  Using the
general method to obtain the encoding circuit for a CWS code
\cite{CWS07} together with some optimization, for the $((5,5,2))$
code we get the circuit shown in Figure~\ref{figencode}.

\begin{figure}
\includegraphics{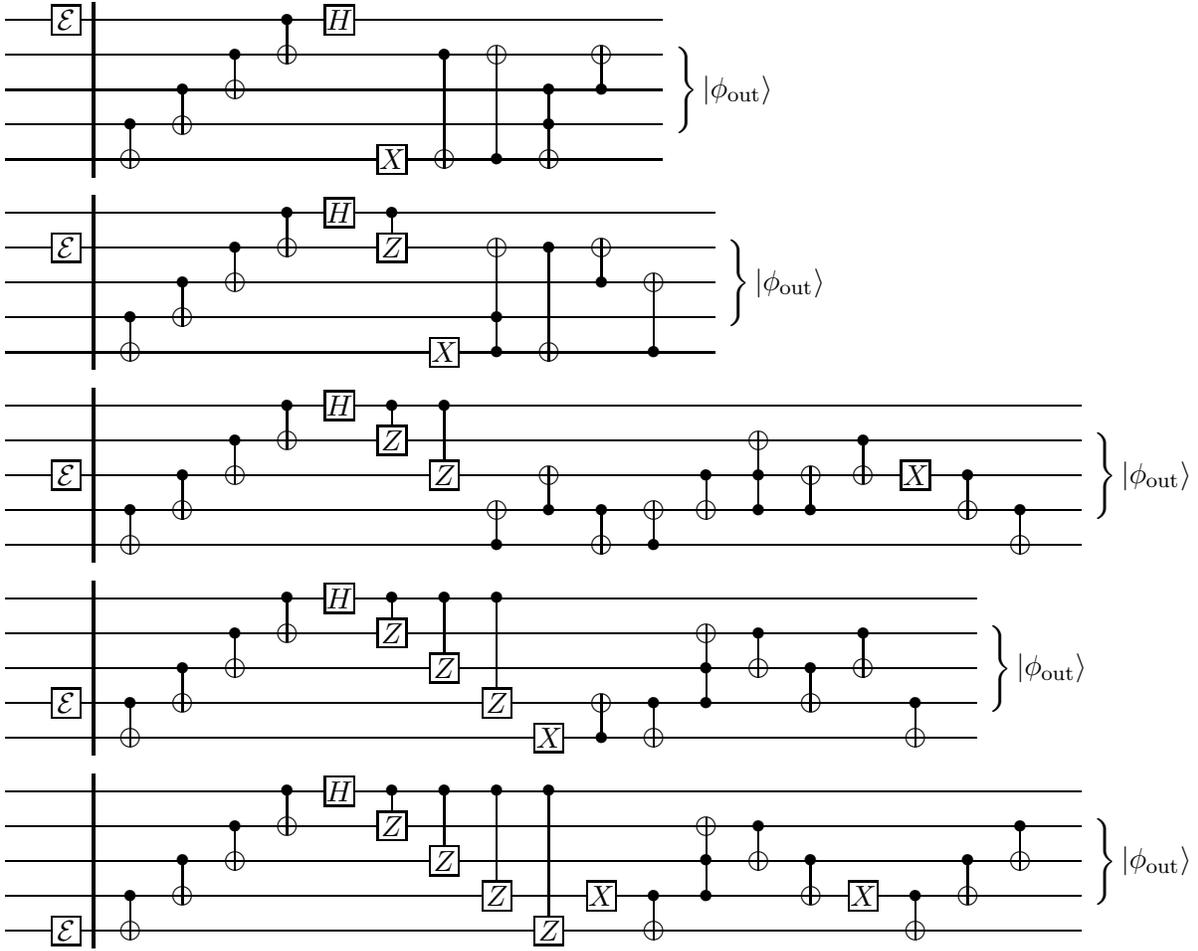}
\medskip
\caption{From top to bottom, the quantum circuits on the right side of
  the double vertical lines are for decoding and correcting the errors
  happening on qubits $1$--$5$, respectively.  $\mathcal{E}$ on the
  left side of the double vertical lines denotes the error
  operation.}\label{figdecode}
\end{figure}

For each location of the error (on one of the five
qubits), a different decoding circuit is needed.
These five decoding circuits are shown in Figure~\ref{figdecode},
for correcting errors happening on qubits $1$--$5$, from top to
bottom, respectively. The error operations, denoted as
$\mathcal{E}$, are also shown in the decoding circuits, separated
by the double vertical lines, noting that $\mathcal{E}$ can be an
arbitrary unitary as well as non-unitary operation.

We consider arbitrary unitary errors which can be represented as
\begin{equation}\label{rot1}
  \mathcal{E} = e^{i\alpha}R_{\hat{n}}(\theta).
\end{equation}
Here $R_{\hat{n}}(\theta) =
e^{-i\theta\hat{n}\cdot\vec{\sigma}/2}$ denoting a rotation by
$\theta$ along the $\hat{n}$ axis, where $\vec{\sigma}$ denotes a
vector with three components formed by the Pauli matrices $X$, $Y$
and $Z$.

When the input state is chosen as
\begin{equation}
\label{inputb}
 |\psi_{b}\rangle = |0\rangle_1|b\rangle_{234}|0\rangle_5,
\end{equation}
the output state after decoding has the form
\begin{equation}\label{outbas}
|\Psi_{b}\rangle = |\Phi\rangle_{15}|b\rangle_{234},
\end{equation}
which recovers the input state $|b\rangle$ of qubits $2$--$4$.  The
syndrome state $|\Phi\rangle_{15}$ contains  information on the error.

Expanding the unitary $\mathcal{E}$ given in Eq. (\ref{rot1}) as a
linear combination of the matrices $E$, $X$, $Z$, and $Y$ gives
\begin{equation}
\label{expE}
 \mathcal{E}=c_{00}E+c_{01}X+c_{10}Z+c_{11}Y.
\end{equation}
Here $E$ is the identity operator acting on a qubit,
and
\begin{alignat*}{5}
c_{00}&{} = e^{i\alpha}\cos(\theta/2),&
c_{01}&{} = -ie^{i\alpha}\sin(\theta/2)n_x,\\
c_{10}&{} = -ie^{i\alpha}\sin(\theta/2)n_z,\quad\text{and}\quad&
c_{11}&{} = -ie^{i\alpha}\sin(\theta/2)n_y, 
\end{alignat*}
where $\hat{n}=(n_x, n_x, n_z)$.  We then obtain the syndrome part
$|\Phi\rangle_{15}$ as
\begin{equation}\label{eqsyn}
 |\Phi\rangle_{15}=c_{00}|00\rangle+c_{01}|01\rangle+c_{10}|10\rangle+c_{11}|11\rangle.
\end{equation}

\section{Experimental Protocol}
\begin{figure}
\includegraphics[width=6in]{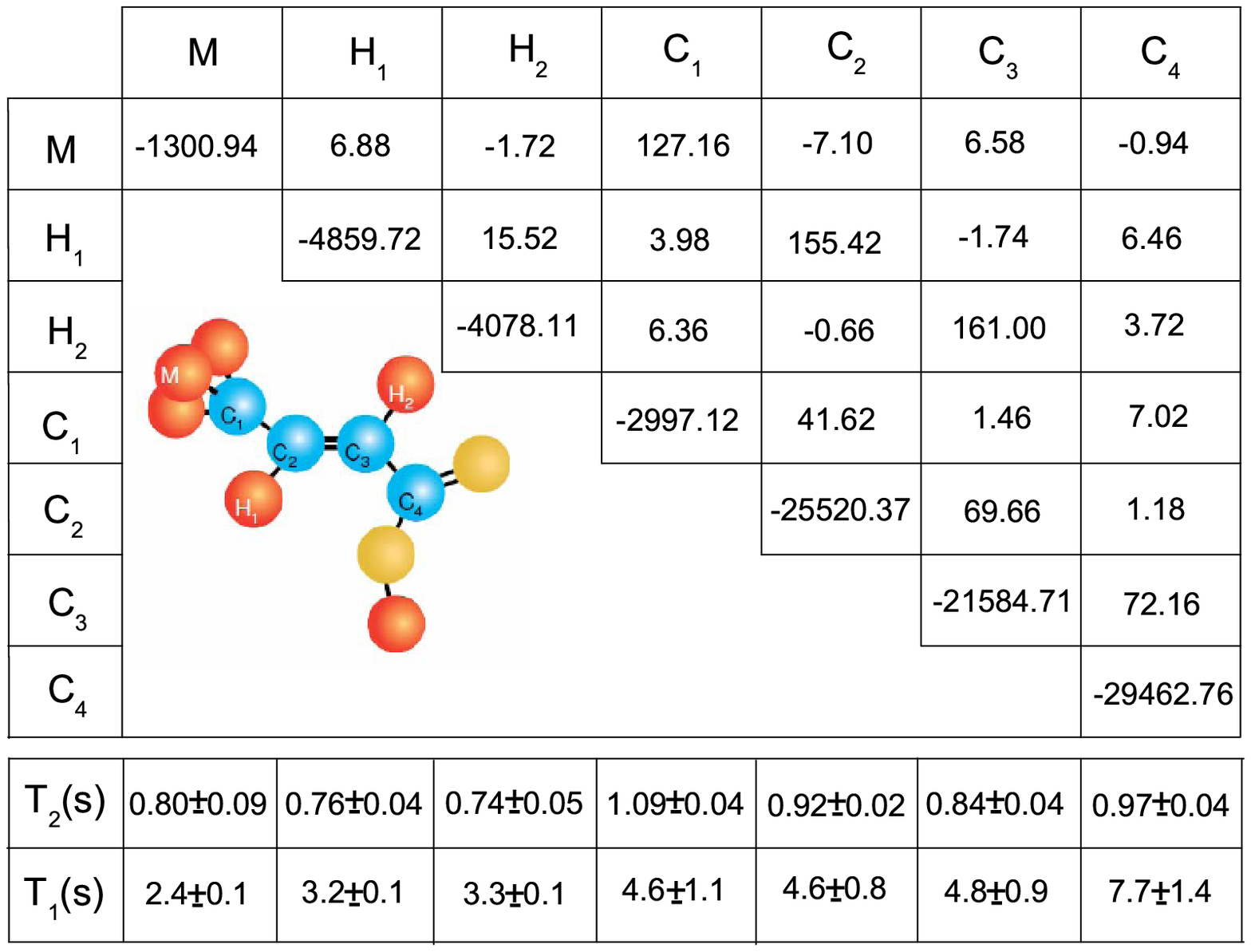}
\caption{(Color online) Characteristics of the molecule of
trans-crotonic acid. The molecular structure is shown as the
inset. The chemical shifts and $J$-coupling constants (in Hz) are
on and above the diagonal in the table. The longitudinal and
transversal relaxation times $T_1$ and $T_2$, which are listed at
the bottom, are estimated by the standard inversion recovery and
Hahn echo sequences. The chemical shifts are given with respect to
reference frequencies of 700.13 MHz (protons) and 176.05 MHz
(carbons). The nine weakly coupled spin half nuclei can provide
seven qubits since the methyl group can be treated as a single
qubit using a gradient-based subspace selection \cite{knill}.}
 \label{figmol}
\end{figure}

Our experiment uses a Bruker DRX 700 MHz spectrometer. We choose
$^{13}$C-labelled trans-crotonic acid dissolved in d6-acetone as
the qubit system \cite{knill,magic}.  The structure of
the molecule is shown as the inset in Figure \ref{figmol}. The
methyl group (denoted as M) can be treated as a spin half nucleus
using a gradient-based subspace selection \cite{knill}. The
Hamiltonian of the seven spins can be represented as ($\hbar = 1$)
\begin{equation}\label{Hnmr}
   H_{NMR} = \sum_{i}\pi \nu_{i} Z_{i} + \sum_{i<j} \frac{\pi}{2} J_{ij} Z_{i}Z_{j}
\end{equation}
where $\nu_{i}$ denotes the chemical shift of spin $i$, and $J_{ij}$ denotes
the scalar coupling strength between spins $i$ and $j$.  The values of
$\nu_{i}$, $J_{ij}$, and relaxation times are listed in Figure
\ref{figmol}. We exploit the seven spins as seven qubits in
experiment.

 We prepare a labelled pseudopure
state $\rho_{s}=\mathbf{0}\mathbf{0}Z
\mathbf{0}\mathbf{0}\mathbf{0}\mathbf{0}$ using the method in Ref.
\cite{knill}. Here the labelled qubit is in state $Z$ and the
other qubits are in state $\mathbf{0}=|0\rangle\langle0|$. From
left to right, the order of the spins is M, H$_1$, H$_2$, C$_1$,
C$_2$, C$_3$, C$_4$. One should note that we are using the
deviation density matrix formalism \cite{ChuangRoy}. In experiment
we do not use spins H$_1$ and H$_2$ in implementing the QECC,
after the preparation of the state $\rho_{s}$. These two spins are
only affected by the hard refocusing pulses. We therefore
concentrate on the subsystem of the five spins M, C$_1$--C$_4$,
denoted as qubits $1$--$5$, in which the QECC is implemented.
Using some basic results on quantum circuits
\cite{PRA95,ChuangAPL}, we transform the circuits for implementing
the QECC shown in Figures \ref{figencode} and \ref{figdecode} into
pulse sequences, shown in Figures \ref{seqsencode} and
\ref{seqsdecode} in the Appendix for encoding and decoding,
respectively. The single-spin rotations along $z$-axis are
implemented through the evolution of the chemical shifts in the
Hamiltonian of the spin system \cite{Grape2}. Rotations along $x$-
and $y$-axes are implemented by standard Isech-shaped r.f.{}
pulses \cite{Grape2,isech1,isech2} for spins M and C$_1$, and
pulses generated by the gradient ascent pulse engineering (GRAPE)
algorithm \cite{Grape1} for C$_2$--C$_4$. The evolutions of the
$J$-couplings between the neighboring qubits are implemented by
numerically optimized refocusing pulses, which are standard Isech
and Hermite-shaped r.f.{} pulses \cite{Grape2,Hermite} and are not
shown here. We choose artificial errors
\cite{RayPRL01,Nature04ion} to demonstrate the QECC, i.e., the
error operations $\mathcal{E}$ in Eq. (\ref{rot1}) are generated
by r.f.{} pulses.  All operations are combined in a custom-built
software compiler to minimize error accumulation
\cite{Grape2,magic}.

We protect the quantum coherence to demonstrate the QECC by putting
one qubit of the input state in a superposition $|+\rangle \equiv
(|0\rangle + |1\rangle)/\sqrt{2}$. The corresponding coherence can be
observed directly in the spectra of the input state and the final
state after the completion of QECC (see below). We avoid the full
quantum state tomography, which would make the demonstration more
difficult because of the required additional experiments and the
limitation of the spectral resolution of certain spins. We choose
three input states as
\begin{equation}
\label{inputs}
   |\psi_{k}\rangle = |0\rangle|s_{k}\rangle|0\rangle
\end{equation}
with
\begin{equation}\label{inputcoh}
 |s_{1}\rangle =|+\rangle|00\rangle,\quad
 |s_{2}\rangle = |01\rangle|+\rangle,\quad
 |s_{3}\rangle = |00\rangle|+\rangle,
\end{equation}
noting that all the basis states in Eq. (\ref{basis5}) are included
here. We introduce error operations of $X$-, $Y$- and $Z$-type by
setting $\hat{n}$ in Eq. (\ref{rot1}) along the $x$-, $y$-, and
$z$-axes, respectively, and choose $\alpha=0$.  Using
Eq. (\ref{outbas}) one can obtain the output states represented as
\begin{equation}\label{outsup}
 |\Psi_{k}\rangle = [\cos(\theta/2) |00\rangle_{15} -
i\sin(\theta/2) |jl\rangle_{15}]\otimes|s_{k}\rangle_{234},
\end{equation}
where $|jl\rangle = |01\rangle$, $|10\rangle$, and $|11\rangle$ for
$X$-, $Z$-, and $Y$-type errors, respectively. Obviously one can
obtain the results for the case of no error ($E$-type error) by
setting $\theta = 0$ in the above equation.

In NMR experiments, we can detect the states of the qubits through
measuring the bulk magnetization \cite{spindynamics,LARay}
\begin{equation}\label{FIDnmr}
 M(t) \propto {\rm Tr}[\rho(t)\sum_{j} (X_{j}+i Y_{j})e^{-t/T_{2,j}^{*}}].
\end{equation}
Here $\rho(t) = U(t) \rho(0) U^{\dagger}(t)$ where $U(t) =
e^{-itH_{NMR}}$, and $\rho(0)$ denotes the density matrix of the state
to be measured after the r.f.{} pulses are switched off.
$T_{2,j}^{*}$ denotes the effective transverse relaxation time of spin
$j$. One can obtain the NMR spectrum through the Fourier transform of
$M(t)$. From some calculations, one can find that the single coherence
elements in $\rho(0)$ can be directly measured in the spectrum, and
the peaks of a certain spin are associated with the states of the
other spins because of the $J$-couplings.

   We first take spectra of C$_1$ for input $|\psi_{1}\rangle$, and
spectra of C${_3}$ for $|\psi_{2}\rangle$ and $|\psi_{3}\rangle$,
respectively. There is only one peak in each spectrum, coming from
the state $|0\rangle|+\rangle|000\rangle$,
$|001\rangle|+\rangle|0\rangle$, or $|000|+\rangle|0\rangle$, in
the input state $|\psi_{1}\rangle$, $|\psi_{2}\rangle$, or
$|\psi_{3}\rangle$, respectively, illustrated by the spectrum
shown as the solid curve in Figure \ref{figpauli} (a) obtained
from $|\psi_{2}\rangle$. The amplitude of the peak is proportional
to the element of the quantum coherence to be protected in state
$|s_{k}\rangle$ in Eq. (\ref{inputcoh}) against the error
operations $\mathcal{E}$. We take the amplitude of the peak in the
spectrum obtained from the input state as the reference to
normalize the corresponding output signals after completion of the
QECC procedure.

In principle, full quantum process tomography (QPT) is required to
completely evaluate the implementation of the QECC and analyze
experimental errors.  Based on the result of the QPT, one can estimate
the experimental errors of the quantum gates for implementing the QECC
through proper models for relaxation effects \cite{bookChuang}. In
practice, however, this task is rather difficult because the number of
experiments required for the QPT increases exponentially with the
involved qubits. In our five qubit case, more than one million
($4^{2n}-4^n=4^{10}-4^{5}$) real numbers have to be determined
\cite{SEQPT11}.  Moreover, the limitation of the spectral resolution
of certain spins in our molecule further increases the difficulty.
More recent strategies for characterizing quantum processes
\cite{SEQPT11,PCQD} might allow an improved evaluation of our QEC
procedure.  Here we choose a simplified approach to demonstrate
robustness of the scheme against artificially introduced errors, and
estimate experimental errors in combination with simulation results.

Noting that in Eq. (\ref{outsup}) the two syndrome qubits are not
entangled with the register qubits, one can extract the state of
the register by tracing over the syndrome qubits in
$|\Psi_{k}\rangle\langle\Psi_{k}|$.  In experiment we only take
the spectra of the register qubits, and implement the partial
trace by adding the splitting of the peaks of the register qubits
caused the coupling of the syndrome qubits. Through Eq.
(\ref{outsup}), one finds that the state
$|0\rangle|+\rangle|000\rangle$, $|001\rangle|+\rangle|0\rangle$,
or $|000\rangle|+\rangle|0\rangle$ contributes a peak with
amplitude
\begin{eqnarray}\label{out00}
   A_{0} &=& \cos^{2}(\theta/2),
\end{eqnarray}
 and
$|j\rangle|+\rangle|00l\rangle$, $|j\rangle|01\rangle|+\rangle
|l\rangle$, or $|j00\rangle|+\rangle |l\rangle$ contributes a peak
with amplitude
\begin{eqnarray}\label{out11}
   A_{1} &=& \sin^{2}(\theta/2)
\end{eqnarray}
in the state $|\Psi_{1}\rangle$, $|\Psi_{2}\rangle$, or
$|\Psi_{3}\rangle$, respectively, after the completion of the QECC,
where the sum $A_{0}+ A_{1} = 1$ shows the robustness against the
error operations. We can therefore benchmark the QECC procedure
through measuring $A_{0}$ and $A_{1}$.
Moreover one can obtain the rotation angle $\theta$ through $A_{0}$
and $A_{1}$ up to its sign. One should notice that we do not measure
the syndrome qubits directly.  The NMR spectrum of the output state is
not necessary exactly in phase with the reference spectrum, due to
experimental errors, e.g. imperfection of r.f.{} pulses. We therefore
choose the absolute values $I_{0} = |A_{0}|$, $I_{1} = |A_{1}|$,
and $I = |A_{0} + A_{1}|$ to represent the results.

\section{Experimental results}

We summarize our experimental results in this section. We have
implemented the $((5,5,2))$ code in three different situations,
for different input states and artificial errors.

Setting A. A single input state $|\psi_{2}\rangle$ as given in Eq.
(\ref{inputs}), and Pauli errors $X$, $Y$, $Z$, and $E$ on each of
the five qubits. The results are summarized in subsection A.

Setting B. A single input state $|\psi_{2}\rangle$ as given in
Eq. (\ref{inputs}), and arbitrary $X$-, $Y$-, and $Z$-type errors,
i.e., arbitrary rotations by an angle $\theta$ along the $x$-, $y$-,
and $z$-axes. We analyze the information on the rotation angle
$\theta$ and sumamrize the results in subsection B.

Setting C. Three input states
$|\psi_{1}\rangle,|\psi_{2}\rangle,|\psi_{3}\rangle$ as given in
Eq. (\ref{inputs}), and arbitrary rotation errors along the
$y$-axis. The results are summarized in subsection C.

For all three situations, our results clearly demonstrate quantum
coherence for implementing the QECC procedure.

\subsection{A. Correcting Pauli errors}
\begin{figure}
\includegraphics[width=4.8in]{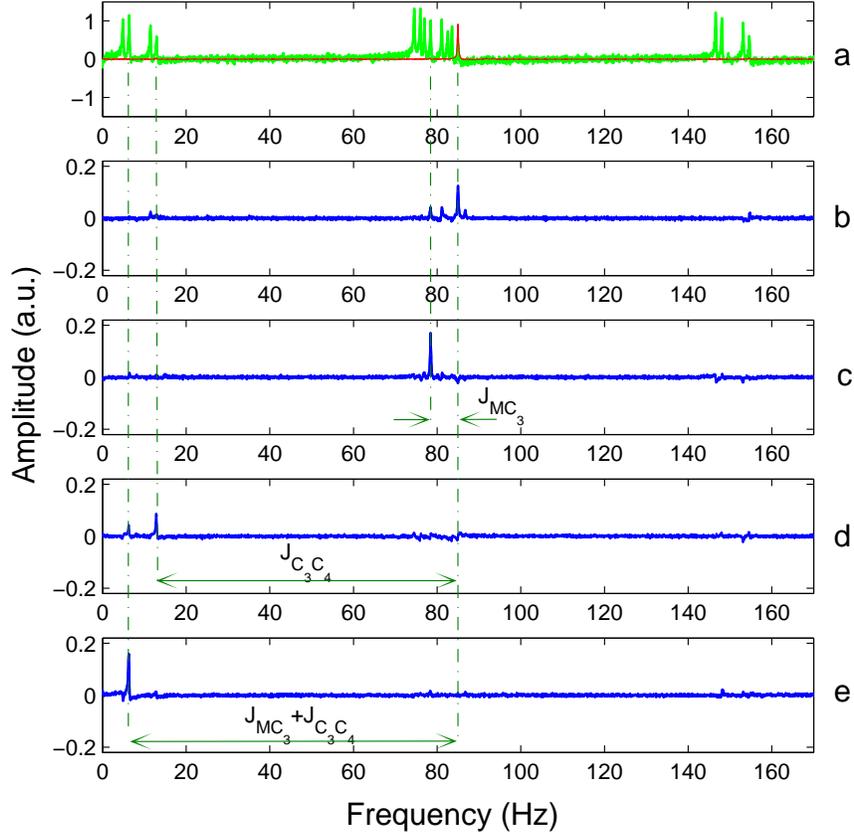}
\caption{(Color online) Experimental NMR spectra of C$_3$ to
  illustrate the results for correcting Pauli errors. 
  In figure (a) the spectrum shown as the
  thin curve is 
   obtained from the input state
  $|\psi_{2}\rangle = |0\rangle|01\rangle|+\rangle|0\rangle$. The
  spectrum shown as the thick curve is obtained from the state of
  the equal weight superposition of all the computational basis
  states, with an applied scale
  factor of 16 for better
  visualization. 
  Figures (b)--(e) illustrate the results of error
  correction for $E$, $Z$, $X$, and $Y$ errors, respectively. The
  shifts of the main peaks, which are indicated
  by the arrows, characterize the type of errors.}
 \label{figpauli}
\end{figure}

Figure \ref{figpauli} shows the experimental NMR spectra of C$_3$
to illustrate the results for correcting Pauli errors, i.e., $E$,
$Z$, $X$, and $Y$ errors. In Figure \ref{figpauli} (a) the
spectrum shown as the thin curve is the reference spectrum,
obtained from the input state $|\psi_{2}\rangle =
|0\rangle|01\rangle|+\rangle|0\rangle$.  The spectrum shown as the
thick curve is obtained from the state of the equal weight
superposition of all the computational basis states (i.e.
$|+\rangle|+\rangle|+\rangle|+\rangle|+\rangle$), where a scale
factor of 16 is applied for better visualization. One can exploit
the positions of the peaks in this spectrum for facilitating to
locate the peaks in other spectra. Figures \ref{figpauli} (b)--(e)
illustrate the results of error correction for $E$, $Z$, $X$, and
$Y$ errors, respectively. The shifts of the main peaks
characterize the type of errors. The distance of the shifts,
measured in $J$-couplings, is indicated by the arrows.

\subsection{B. Correcting $X$-, $Y$-, and $Z$-type errors}
\begin{figure}
\includegraphics[width=4.8in]{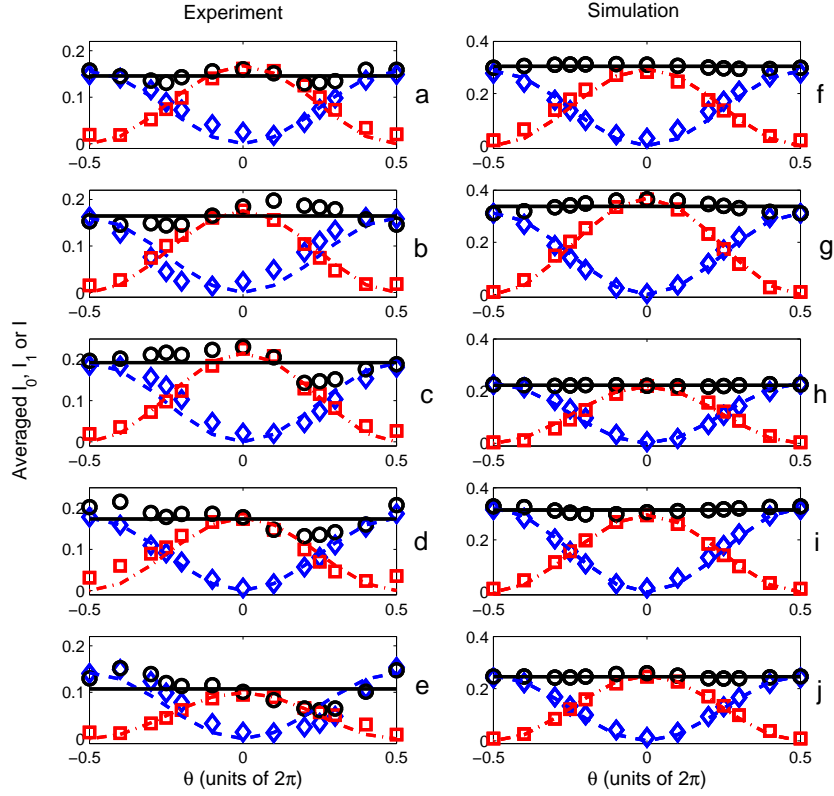} 
\caption{(Color online) Averaged results of the correction for $X$-,
  $Y$- and $Z$-type errors. The input state is chosen as
  $|\psi_{2}\rangle$, and the signal in the input state is chosen as
  reference to normalize the signals after the completion of the QECC
  procedure. The left and right columns show the results in experiment
  and by simulation, respectively.  The results for errors on qubits
  $1$--$5$ are shown from top to bottom, respectively. The averaged
  $I_{0}$ and $I_{1}$, denoted as $\bar{I}_{0}$ and $\bar{I}_{1}$, are
  indicated by squares and diamonds, respectively. In comparison with
  the theoretical values $\bar{I}_{0}^{th}$ and $\bar{I}_{1}^{th}$,
  the data can be fitted as $\alpha_{0}\bar{I}_{0}^{th}$ and
  $\alpha_{1}\bar{I}_{1}^{th}$ shown as the dash-dotted and dashed
  curves in figures (a)--(e) for experiment and (f)--(j) for
  simulation, where $\alpha_{0}$ and $\alpha_{1}$ are constant
  coefficients in each figure and listed in the main text.  The
  averaged $I$, denoted as $\bar{I}$ and marked as circles, indicates
  the intensity of the remaining signals after completion of the QECC
  procedure.  The remaining signal is robust against the
  rotation angle $\theta$, demonstrating the success of the QEC
  procedure.}
 \label{figXYZ}
\end{figure}

The input state is chosen as $|\psi_{2}\rangle$. The results for
correcting $X$-, $Y$-, and $Z$-type errors with arbitrary
rotations are shown in Figure \ref{figXYZ}. We exploit the
averaged $I_{0}$, $I_{1}$ and $I$, respectively denoted as
$\bar{I}_{0}$, $\bar{I}_{1}$ and $\bar{I}$, over the results of
correcting the three types of errors as benchmark. The
experimental results for correcting errors on qubits $1$--$5$ are
shown as Figures \ref{figXYZ} (a)--(e), respectively, where
$\bar{I}_{0}$, $\bar{I}_{1}$, and $\bar{I}$ are indicated by
squares, diamonds and circles. In comparison with the theoretical
values $\bar{I}_{0}^{th}$ and $\bar{I}_{1}^{th}$, the experimental
data can be fitted as $\alpha_{0}\bar{I}_{0}^{th}$ and
$\alpha_{1}\bar{I}_{1}^{th}$ shown as the dash-dotted and dashed
curves. We fit the data for $\bar{I}$ using a constant function,
shown as the plane lines. The remaining quantum coherence is
robust against the rotation angle $\theta$, demonstrating the
success of the QECC procedure. The values of the fitted
$\alpha_0$, $\alpha_1$ and $\bar{I}$ are listed in Table
\ref{tableB}. We also list the results by simulation in Figures
\ref{figXYZ} (f)--(j), for correcting the errors happening on
qubits $1$--$5$, respectively. In simulation, we take into account
the $T_2$ effects, imperfection in refocusing protocols and the
theoretical infidelity of the numerically generated pulses.
 The fitting results for  $\alpha_{0}$, $\alpha_{1}$
and $\bar{I}$ are also listed in Table \ref{tableB}.

\begin{table} 
\begin{tabular}{|c|c|c|c|c|c|}
\hline Error location &  1   &   2 & 3 & 4 & 5  \\
\hline $\alpha_0$ (experiment) &  $0.156 \pm 0.020$   &   $0.160
\pm 0.038$ & $0.191 \pm 0.036$ & $0.177 \pm 0.010$ & $0.142 \pm 0.038$  \\
\hline $\alpha_1$ (experiment) &  $0.168 \pm 0.023$   &   $0.176
\pm 0.019$ & $0.215 \pm
0.025$ & $0.177 \pm 0.038$ & $0.099 \pm 0.016$  \\
\hline $\bar{I}$ (experiment) &  $0.146 \pm 0.012$   &   $0.164
\pm 0.018$ &
$0.193 \pm 0.028$ & $0.174 \pm 0.027$ & $0.107 \pm 0.029$  \\
\hline $\alpha_0$ (simulation) &  $0.288 \pm 0.031$   &   $0.312
\pm 0.018$ & $0.230 \pm 0.016$ & $0.322
\pm 0.019$ & $0.249 \pm 0.015$  \\
\hline $\alpha_1$ (simulation) &  $0.295 \pm 0.037$   &  $0.369
\pm 0.022$ & $0.218 \pm 0.020$ & $0.295 \pm 0.016$ & $0.250
\pm 0.013$  \\
\hline $\bar{I}$ (simulation) &  $0.304 \pm 0.007$   &   $0.337
\pm 0.018$ &
$0.222 \pm 0.003$ & $0.315 \pm 0.010$ & $0.246 \pm 0.006$  \\
\hline
\end{tabular}
\caption{ The fitted  $\alpha_0$, $\alpha_1$ and $\bar{I}$ in
experiment and by simulation in case B. }\label{tableB}
\end{table}

\subsection{C. Correcting $Y$-type errors from various input states}

\begin{figure}
\includegraphics[width=5in]{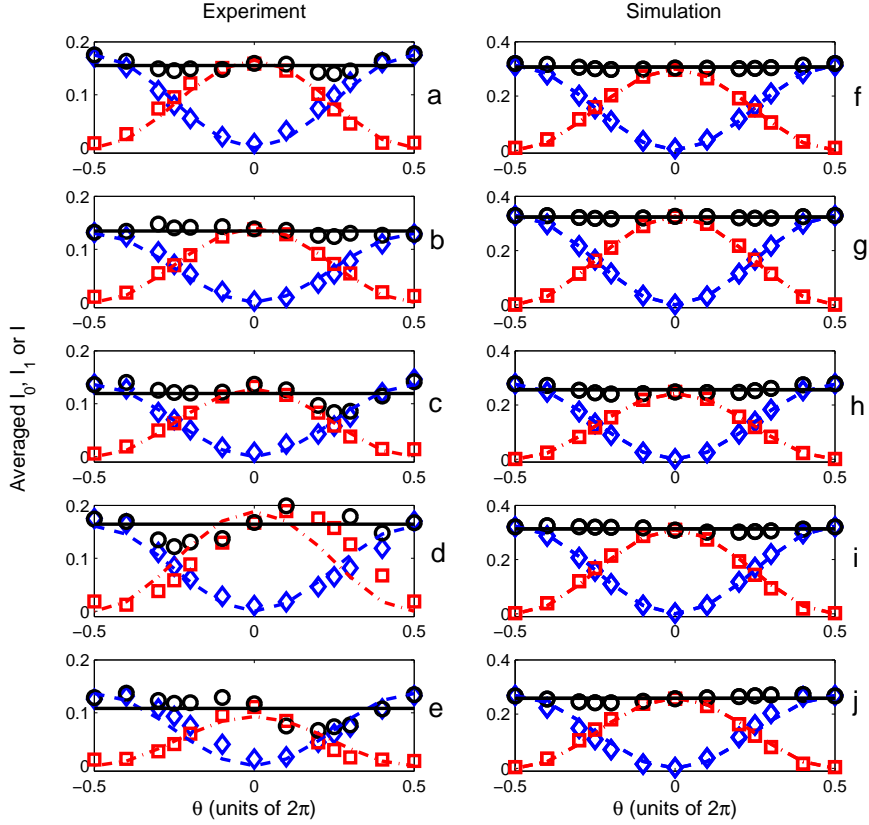} 
\caption{(Color online) Averaged results over various input states
  $|\psi_{k}\rangle$ ($k = 1, 2, 3$) for correcting $Y$-type
  errors. The left and right columns show the results in experiment
  and by simulation, respectively. The results for errors on qubits
  $1$--$5$ are shown from top to bottom, respectively. We choose the
  signals of the input states as reference to normalize the signals
  after the completion of the QECC procedure. $\bar{I}_{0}$ and
  $\bar{I}_{1}$ are indicated by squares and diamonds,
  respectively. In comparison with the theoretical values
  $\bar{I}_{0}^{th}$ and $\bar{I}_{1}^{th}$, the data can be fitted as
  $\alpha_{0}\bar{I}_{0}^{th}$ and $\alpha_{1}\bar{I}_{1}^{th}$ shown
  as the dash-dotted and dashed curves in figures (a)--(e) for
  experiment and (f)--(j) for simulation, where $\alpha_{0}$ and
  $\alpha_{1}$ are listed in the main text.  The intensity $\bar{I}$
  of the remaining signal, marked as circles, is robust against the
  rotation angle $\theta$, demonstrating the success of the QEC
  procedure. }
 \label{figexp}
\end{figure}

 We measure $I_{0}$, $I_{1}$, and $I$ for the three
input states $|\psi_{k}\rangle$ with $k = 1$, $2$, $3$, and
exploit the average over the input states, respectively denoted as
$\bar{I}_{0}$, $\bar{I}_{1}$ and $\bar{I}$, to represent the
results of the error correction. The results in the experimental
implementation and by simulation are shown in Figures \ref{figexp}
(a)--(e), and (f)--(j), corresponding to errors happening on
qubits $1$--$5$, respectively, where  $\bar{I}_{0}$, $\bar{I}_{1}$
and $\bar{I}$ are indicated by squares, diamonds and circles.  In
comparison with the theoretical values $\bar{I}_{0}^{th}$ and
$\bar{I}_{1}^{th}$, the experimental data can be fitted as
$\alpha_{0}\bar{I}_{0}^{th}$ and $\alpha_{1}\bar{I}_{1}^{th}$
shown as the dash-dotted and dashed curves in Figures \ref{figexp}
(a)--(e). The data for $\bar{I}$ can be fitted as a constant
function, and the fitting results are shown as the plane lines.
The remaining quantum coherence is robust against the rotation
angle $\theta$, demonstrating the success of the QECC procedure.
The values of the fitted $\alpha_0$, $\alpha_1$ and $\bar{I}$ for
the results in experiment and by simulation are listed in Table
\ref{tableC}.

\begin{table} 
\begin{tabular}{|c|c|c|c|c|c|}
\hline Error location &  1   &   2 & 3 & 4 & 5  \\
\hline $\alpha_0$ (experiment) &  $0.175 \pm 0.013$   &   $0.130
\pm 0.011$ & $0.136 \pm
0.012$ & $0.161 \pm 0.022$ & $0.137 \pm 0.026$  \\
\hline $\alpha_1$ (experiment) &  $0.165 \pm 0.017$   &   $0.140
\pm 0.010$ & $0.128 \pm 0.010$ & $0.188 \pm 0.063$ & $0.093 \pm 0.019$  \\
\hline $\bar{I}$ (experiment) &  $0.155 \pm 0.012$   &   $0.134
\pm 0.007$ & $0.119 \pm 0.018$ & $0.165 \pm
0.027$ & $0.108 \pm 0.024$  \\
\hline $\alpha_0$ (simulation) &  $0.314 \pm 0.008$   &   $0.330
\pm 0.002$ & $0.276 \pm
0.005$ & $0.324 \pm 0.008$ & $0.269 \pm 0.031$  \\
\hline $\alpha_1$ (simulation) &  $0.301 \pm 0.012$   &  $0.327
\pm 0.005$ & $0.243 \pm 0.005$ & $0.311 \pm 0.017$ &
$0.260 \pm 0.016$  \\
\hline $\bar{I}$ (simulation) &  $0.307 \pm 0.008$   &   $0.323
\pm 0.004$ & $0.257 \pm 0.014$ & $0.314
\pm 0.008$ & $0.258 \pm 0.011$  \\
\hline
\end{tabular}
\caption{ The fitted  $\alpha_0$, $\alpha_1$ and $\bar{I}$ in
experiment and by simulation in case C.}\label{tableC}
\end{table}

\section{Discussion of the experimental results}
\subsection{A. Extracting partial information of error operations}
\begin{figure}
\includegraphics[width=6in]{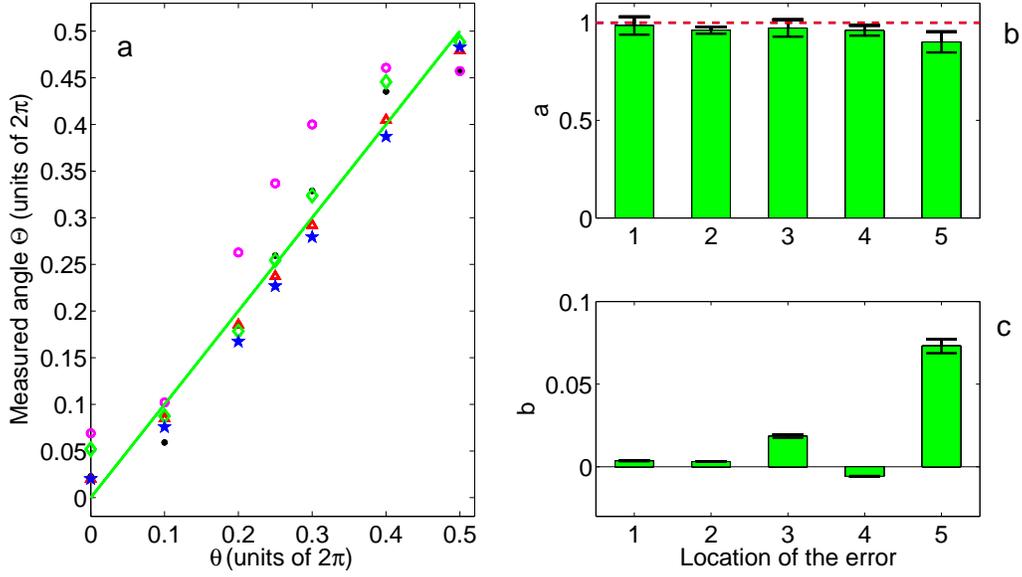} 
\caption{(Color online) Experimental results of error syndromes.
  Using the measured $\bar{I}_{0}$ and $\bar{I}_{1}$, we obtain the
  absolute values of the rotation angles for the errors, denoted by
  $\Theta$.  The theoretical expectation $\Theta = \theta$ is shown as
  the line. The experimental data for errors happening on qubits
  $1$--$5$ are denoted as filled circles, triangles, diamonds, stars,
  and empty circles, respectively.  The deviation of the data for
  errors happening at qubit 5 is larger than the other cases, because
  more r.f.{} pulses are required for correcting the errors at qubit 5
  (see Figure \ref{seqsdecode}).  We exploit $\Theta = a\theta + b$ to
  fit the data. The coefficients $a$ and $b$ are shown in figure (b)
  and (c), respectively, where the dashed line indicates the
  theoretical value $a = 1$.}
 \label{fignumdur}
\end{figure}

Through measuring the qubits of the register [see Eqs. (\ref{out00})
  and (\ref{out11})], we can obtain partial information of the errors,
i.e., the types of the errors illustrated in Figure
\ref{figpauli}, and the absolute values of the rotation angles of
the errors, exploiting the couplings between the register and
syndrome qubits. Figure \ref{fignumdur} illustrates the
experimental results obtained from the data in Figure \ref{figexp}
for extracting absolute values of the rotation angles. In Figure
\ref{fignumdur} (a), the theoretical expectation $\Theta = \theta$
is shown as the line. The experimental data is distributed near
the line. We fit the data using a linear function as $\Theta =
a\theta + b$. The coefficients $a$ and $b$ are shown in Figure
\ref{fignumdur} (b) and (c), respectively, where the dashed line
indicates the theoretical value of $a = 1$. The fitting results
show a good agreement between experiment and theory.

\subsection{B. Error sources in implementation}
The simulation is mainly used to analyze experimental errors.
Excluding the preparation of the pseudopure state, the number of
r.f.{} pulses ranges from about $320$ to $440$, and the duration
of the experiments ranges from about $0.55$s to $0.72$s. The
duration is comparable with, and indeed very close to the
transverse relaxation times ($T_{2}$'s in Figure \ref{figmol}).
Hence the limitation of the coherence time contributes major
experimental errors. Through comparing the results by simulation
with and without $T_{2}$ effects, we estimate that the limitation
of $T_{2}$ contributes to about $34\%$--$45\%$ of the loss of
signal. Additionally, the errors due to imperfections of the
refocusing protocols and the implementation of r.f.{} pulses
contribute about $10\%$--$16\%$ and $3\%$--$19\%$ of the loss of
signal, respectively.

\subsection{C. Comparison with the implementation of the $((5,2^1,3))$ code}
We use the same seven-spin-qubit NMR system that was used in the
implementation of the $((5,2^1,3))$ codes \cite{RayPRL01}.  It should
be emphasized that in the $((5,2^1,3))$ experiment, the entire circuit
one would need to carry out consists of four parts: encoding,
(artificial) errors, decoding, and recovery. The decoding circuit is
just the reverse of the encoding circuit, but the recovery circuit
involves approximately three times more gates than the
encoding/decoding circuit. However, because the artificial errors
demonstrated for the code $((5,2^1,3))$ are all Pauli errors, after
the decoding circuit using only Clifford gates (gates that map Pauli
operators to Pauli operators), the five qubits are in a product
state. Only a single qubit carries information that needs a recovery
operation, while the other four qubits carry error syndrome
information. Consequently, the noise (mainly dephasing noise) during
the recovery procedure does not have much effect on the final
experimental fidelity, as coherence is only needed to be maintained on
a single qubit.

On the contrary, we use three qubits to carry information.
Although our experiment requires implementing quantum circuits
some of which are of comparable size as those for the
$((5,2^1,3))$ code, quantum coherence in a larger Hilbert space
has to be maintained throughout the entire procedure of the code.
In implementation, maintaining coherence in a larger space e.g.
three qubits, is more difficult than in a one-qubit space, because
the high order coherence could decay faster than the single
coherence under effects of decoherence \cite{CoryNMRQC}. This case
might explain the low remaining signals in our experiment
($10$--$20 \%$), compared with the $((5,2^1,3))$ experiment where
the intensity of the remaining signals ranges $48$--$87\%$.

\section{Conclusion}
We implement a five-qubit quantum error-correcting code using NMR.
The code protects an arbitrary state in a five-dimensional Hilbert
space, and is capable of correcting a single erasure error. The
code is a CWS code which encodes a larger Hilbert space than any
stabilizer code with the same length and being capable of
correcting a single erasure error. Our results demonstrate a good
nonadditive quantum code in experiment for the first time.

Compared with the previous implementation of another five-qubit code
using the same seven-spin-qubit system~\cite{RayPRL01}, which encodes
a two-dimensional Hilbert space and is capable of correcting an
arbitrary single-qubit error, the pulse sequences in our experiment
are more complex. In order to shorten the length of sequences, we
exploit pulses optimized by the GRAPE algorithm to implement the
$\pi/2$ spin-selective pulses for certain spins. The duration of our
experiments ranges from $0.55$s to $0.72$s, which challenges the
ultimate limit of coherent control of the system, with a typical $T_2$
time ranging from $0.8$s to $1.1$s.  Despite experimental
imperfections which induce signal loss, the signal remains
resilient against the artificial errors.

\section*{Acknowledgement}
We thank Industry Canada for support at the Institute for Quantum
Computing. R.L. and B.Z. acknowledge support from NSERC and CIFAR.
Centre for Quantum Technologies is a Research Centre of Excellence
funded by the Ministry of Education and the National Research
Foundation of Singapore.

\appendix{{\bf Appendix: Pulse sequences}}

\begin{figure}[hbt]
 \includegraphics[width=\hsize]{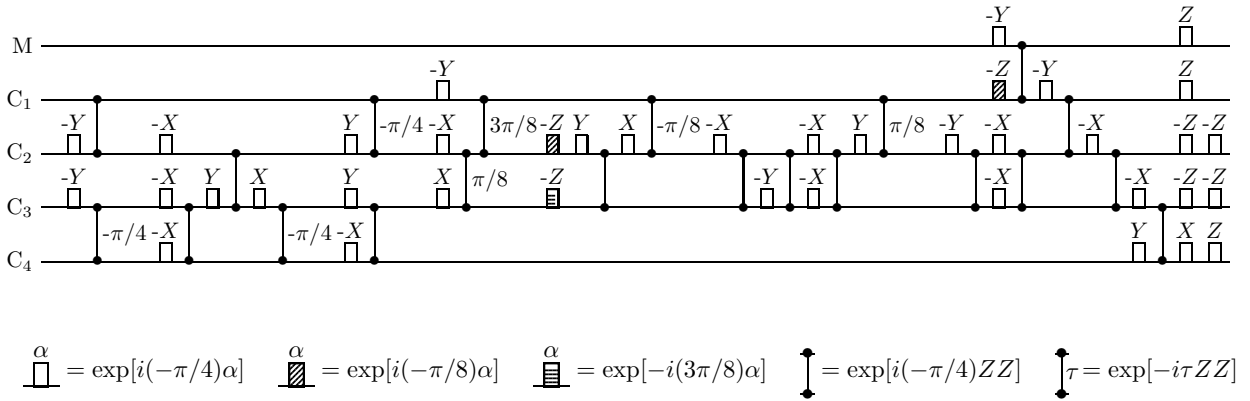}
\caption{Pulse sequences to encode the $((5,5,2))$
  code.}\label{seqsencode}
\end{figure}

\begin{figure}[hbt]
 \includegraphics[width=\hsize]{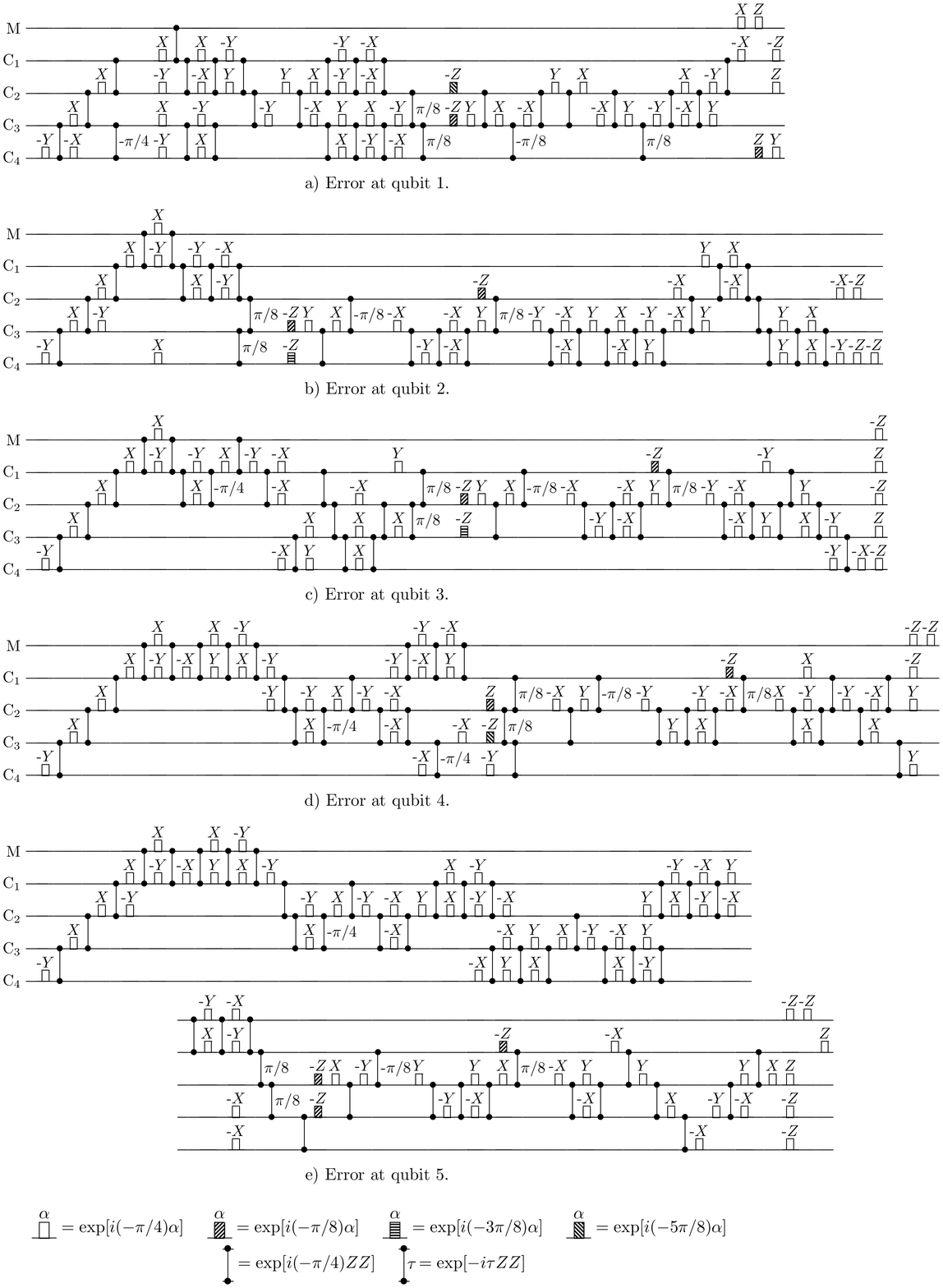}
\caption{Pulse sequences to decode the $((5,5,2))$ code, for errors
  happening on each of the five qubits.}\label{seqsdecode}
\end{figure}

We list the pulse sequences for implementing the encoding and
decoding circuits as Figures \ref{seqsencode} and
\ref{seqsdecode}, respectively.



\begin{thebibliography}{}
\bibitem{note1} In this article, we restrict ourselves to qubit
  systems, but the theory of QECCs can be applied to systems of
  arbitrary dimension.
\end{thebibliography}

\clearpage


\end{document}